\documentclass{emulateapj}

\usepackage{graphics,graphicx}
\usepackage{natbib}
\def\apj{\rm ApJ}
\def\apjl{\rm ApJL}
\def\apjs{\rm ApJS}
\def\aj{\rm AJ}
\def\mnras{\rm MNRAS}
\def\nat{\rm Nature}

\def\pasp{\rm PASP}

\newcommand{\msun}{M$_{\odot}$}

\shorttitle{Wandering BHs in Disk Galaxy Halos}
\shortauthors{Bellovary et al.}

\begin{document}

\title{Wandering Black Holes in Bright Disk Galaxy Halos}

\author{Jillian M. Bellovary\altaffilmark{1}, Fabio
Governato\altaffilmark{1}, Thomas R. Quinn\altaffilmark{1}, James
Wadsley\altaffilmark{2}, Sijing
Shen\altaffilmark{2}, Marta Volonteri \altaffilmark{3}}

\altaffiltext{1}{Department of Astronomy, University of Washington, Seattle, WA} 
\altaffiltext{2}{Department of Physics and Astronomy, McMaster University, Hamilton, ON, Canada}

\altaffiltext{3}{Department Astronomy, University of Michigan, Ann Arbor, MI}

\begin{abstract}

\noindent

We perform SPH+N-body cosmological simulations of massive disk
galaxies, including a formalism for black hole seed formation and
growth, and find that satellite galaxies containing supermassive black
hole seeds are often stripped as they merge with the primary galaxy.
These events naturally create a population of ``wandering'' black
holes that are the remnants of stripped satellite cores; galaxies like
the Milky Way may host 5 -- 15 of these objects within their halos.
The satellites that harbor black hole seeds are comparable to Local
Group dwarf galaxies such as the Small and Large Magellanic Clouds;
these galaxies are promising candidates to host nearby intermediate
mass black holes. Provided that these wandering black holes
retain a gaseous accretion disk from their host dwarf galaxy, they
give a physical explanation for the origin and observed properties
of some recently discovered off-nuclear ultraluminous X-ray sources
such as HLX-1.

\end{abstract}


\section{Introduction}

Recent evidence for the existence of intermediate mass black holes
(IMBHs) raises questions about how such objects might form and evolve.
IMBH candidates exist in globular clusters \citep{Ulvestad07}, nearby
bulgeless galaxies \citep{Filippenko03, Barth04}, and active galactic
nuclei \citep{Greene04}.  Additionally, off-nuclear ultraluminous
X-ray sources (ULXs) have become increasingly promising IMBH
candidates \citep{Farrell09, Jonker10}.  A source of IMBHs may be the
seeds of supermassive black holes (SMBHs) formed at high redshift; any
seed that does not grow into a SMBH would today be observed as an
IMBH.  While the precise mechanism for seed black hole (BH) formation
is unknown, there are several postulated theories.  One possible
mechanism for SMBH seed formation is the direct collapse of pristine,
low angular momentum gas, which forms BHs with mass on the order of
$10^4 - 10^6$ \msun~ \citep{Koushiappas04,Lodato06,Begelman06}.
Another possibility is that the seeds are the remnants of Population
III stars, with masses around $10^2 - 10^3$ \msun~
\citep{Madau01,Volonteri03}.  Alternatively, the first nuclear star
clusters may collapse to form BHs of mass $1000 - 2000$ \msun~
\citep{Devecchi09}.

We perform SPH+$N$-body simulations of massive disk galaxies to
explore the evolution of seed BHs in a cosmological context.  While
previous works have addressed the growth of central SMBHs
\citep{DiMatteo08,Okamoto08,Booth09}, here we specifically focus on
those which do not end up in galaxy centers.  We include BH growth by
gas accretion and merging, as well as radiative feedback in order to
form a fully self-consistent picture of BH growth and evolution.  We
find that the tidal stripping of galaxies containing SMBH seeds leads
to a population of ``wandering'' BHs within the larger galaxy halo.
Such a phenomenon has been predicted in the case of idealized
simulations of merging galaxies \citep{Governato94,Stelios05}, though
fully cosmological simulations in the current $\Lambda$CDM paradigm
are needed to test a suite of non-idealized scenarios.  Wandering BHs
may also be created by a gravitational slingshot from 3-body black
hole interactions \citep{Volonteri05}, gravitational recoil, or a
distribution of Population III star remnants which have not spiralled
into the central BH \citep{Schneider02}.  In each of these cases, the
timescale for dynamical friction is longer than a Hubble time
\citep{Taffoni03}, leaving a potentially substantial population of
``wandering'' IMBHs in galaxy halos.

The recent discovery of an off-nuclear IMBH candidate
\citep{Farrell09} leads us to explore whether such an object can be
explained by a ``wandering'' BH.  The object HLX-1 is offset from the
nucleus of its host spiral galaxy, and exhibits an X-ray luminosity of
$10^{42}$ ergs s$^{-1}$.  This luminosity implies a lower limit to a
black hole (BH) mass of 500 \msun, but the true mass may be much
larger.  We discuss under what circumstances HLX-1 would be
observable.  Preliminary results suggest that HLX-1 cannot be
described by an isolated wandering BH, but a BH traveling with the
remnant core of its parent galaxy may explain its observed properties.

\section{Simulations}

We use the N--body code GASOLINE \citep{Wadsley04,Stadel01}, an SPH
tree code which incorporates star formation, gas cooling and
hydrodynamics, and supernova feedback, and successfully models
realistic galaxies.  Studies of cosmological simulations with GASOLINE
have produced galaxies that follow the mass-metallicity relation
\citep{Brooks07} and the HI Tully-Fisher relation \citep{Governato09},
exhibit cold flow gas accretion \citep{Brooks09}, and reproduce the
distribution of Damped Lyman Alpha systems at high redshift
\citep{Pontzen08}. Additionally, \citet{Governato10} have shown how
high resolution and physically motivated supernova feedback allow for
the formation of a bulgeless dwarf galaxy with a dark matter core.
The ability to form realistic disk galaxies is critical to our
analysis of BH physics, because of the need to accurately trace the
angular momentum of inspiraling gas which comprises the disk and
eventually fuels the BH, as well as the star formation histories
responsible for dispensing metals into the ISM via supernova feedback.

For this Letter, we have simulated four different Milky Way-mass
halos, which were selected from a uniform, 50 Mpc volume and
resimulated at a high resolution using the volume renormalization
technique \citep{Katz93}.  All simulations are run with gravitational
softening $\epsilon = 0.3$ kpc, gas particle masses of $2.28 \times
10^5$ \msun, dark matter particle masses of $1.26 \times 10^5$ \msun,
a Chabrier initial mass function \citep{Chabrier}, and a {\em WMAP}
year 3 cosmology \citep{WMAP3}.  Star formation and supernova recipes
are described in detail in \citet{Stinson06} and \citet{Governato07};
we adopt parameter values $c = 0.1$ and $eSN = 1.0$.  Stars are
allowed to form when gas reaches a threshold density of 1.0 amu
cm$^{-3}$, which allows us the most physical representation of star
formation given our resolution. While our dwarf galaxy satellites are
not resolved to the level of \citet{Governato10}, for our purposes
here this resolution is sufficient.  Simulation properties are
described in Table 1.  Circular velocity $V_{circ}$ is measured using
the width at 20\% of the peak of the simulated HI line profile
\citep[see][]{Governato09}.  The disk scale length $r_s$ is determined
by simultaneously fitting an exponential + Sersic profile to the
projected edge-on stellar surface density of the galactic disk.

\begin{deluxetable}{lcccll}
\tablecolumns{6} 
\tablewidth{0pc}
\tablecaption{Simulation Properties}
\tablehead{
\colhead{Run} & \colhead{\# within R$_{vir}$} &\colhead{M$_{vir}$ } &\colhead{V$_{circ}$ } & \colhead{R$_{vir}$ } & \colhead{$r_s$ }\\
\colhead{}&\colhead{}&\colhead{(\msun)}&\colhead{(km/s)}&\colhead{(kpc)}&\colhead{(kpc)} }

\startdata
h239 & 7408639 & $8.32 \times 10^{11}$ & 247.7 & 242.9 & 2.82\\
h258 & 7347383 & $7.90 \times 10^{11}$ & 225.4 & 238.8 & 3.81\\
h277 & 6624914 & $7.09 \times 10^{11}$ & 262.7 & 230.3 & 2.74\\
h285 & 7373714 & $8.18 \times 10^{11}$ & 238.2 & 241.5 & 2.24\\

\enddata
\end{deluxetable}


There is much uncertainty regarding the formation of the initial
``seed'' black holes which grow to become SMBHs; however, they must
form early on and grow quickly in order to form $z \sim 6$ quasars
\citep{Fan01}.  We maintain a physically motivated scenario by
assuming seed black holes form via the direct collapse of extremely
low metallicity gas as in \citet{Begelman06}.  However, our choice of
seed formation scenario is not crucial. We focus here on the dynamical
evolution of seed BHs, which will be the same regardless of where we
set the initial mass (i.e. 100 \msun~ for the first stars or $10^5$
\msun~ for direct collapse). The dynamical friction timescale is
longer than the Hubble time even for these relatively heavy seeds,
given the local densities involved here. Seeds are allowed to form if
the parent gas particle meets the criteria for star formation
\citep[see][]{Stinson06} and additionally has zero metallicity.  We
designate a probability that a newly formed star will instead become a
seed black hole with mass $M_{BH} = 2.28 \times 10^5 M_\sun$.  This
probablility is set to 0.10 in order to reproduce the predicted black
hole seed halo occupation probability at $z = 3$ \citep{Volonteri08}.
Allowing seed BHs to form out of only pristine gas results in the
truncation of seed BH formation at a redshift of $\sim 3.5$ due to the
efficient diffusion of metals produced by the first supernovae into
the ISM.  This criterion causes seed BHs to form in the regions of
early bursts of star formation, which are primarily the centers of the
first massive halos to form in the simulation. In the event that more
than one BH forms at the same time within a softening length, the BHs
are merged into a single BH particle, with $M_{BH}$ as the total of
the individual masses.  However, we make no assumptions regarding the
large-scale properties of the host halo, allowing the BHs to form
based only on the properties of the local environment.  BHs are not
fixed at their halo centers, but are allowed to move as the forces
upon them dictate.  We minimize the effects of two-body interactions
by adopting dark matter particle masses that are similar to gas
particle masses in the high resolution region, which greatly helps the
central BHs stay at their halo centers.

Black holes are presumed to accrete gas isotropically, following the
formula for Bondi--Hoyle accretion.  The accretion rate is limited by
the Eddington rate, assuming a 10\% radiative efficiency.  In
addition, a fraction of the rest mass energy of the accreted gas is
converted to radiative energy, which is then isotropically imparted
onto the surrounding gas \citep{DiMatteo05}.  To dissipate the
feedback energy in a realistic way, we use a blast-wave feedback
approach similar to the supernova feedback recipe described in
\citet{Stinson06}.  We set the fraction of radiated energy given to
the surrounding gas to be 0.1\%.

Black holes are allowed to merge if they {\em (a)} are within one
another's softening length and {\em (b)} fulfill the criterion
$\frac{1}{2} \Delta \vec{v}^2 < \Delta \vec{a} \cdot \Delta \vec{r}$,
where $\Delta \vec{v}$ and $\Delta \vec{a}$ are the differences in
velocity and acceleration of the two BHs, and $\Delta \vec{r}$ is the
distance between them.

We identify galaxies and their halos with AHF \citep{Knebe01,Gill04},
which identifies a virial radius based on the overdensity criterion
for a flat universe \citep{Gross97}.  For each output, we identify
every halo and all of the particles it contains; thus, we can track a
BH's halo residency at each timestep.  We use this process to define
whether a black hole is stripped from its parent halo to reside in
another.  A BH is denoted as ``stripped'' if its parent halo
overdensity can no longer be detected as a separate halo.  The BH is
then identified as a resident of the larger galaxy, most likely in the
outer regions.

\section{Results}

We provide a dynamical mechanism to place massive BHs in galaxy
halos, which is simply a consequence of hierarchical merging in the
$\Lambda$CDM paradigm.  Detecting these objects may help constrain the
mechanism of seed BH formation, and we investigate a few scenarios in
which a wandering BH may be observed as a ULX.

\subsection{Wandering Black Holes}

\begin{figure}[htb]
\begin{center}
\plotone{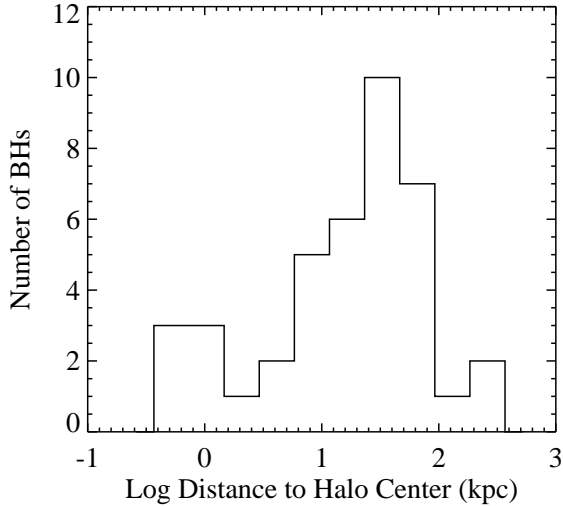}

\caption{{\em Top Panel:} Distribution of black hole radial distances
to their halo centers for 40 black holes in four simulated galaxies.
\label{fig:radial}}

\end{center}
\end{figure}

For our four simulated halos, we show the radial distribution of BHs
at $z = 0$, defined as the distance between the halo center and the
BH, in Figure \ref{fig:radial}.  Each of the four galaxies has a
central BH and between 5 and 15 ``wandering'' BHs whose distances
range from 10 - 100 kpc.  The vast majority of wandering BHs have
grown in mass by less than 2\% since their formation.  A small number
(5 out of 36 total) have undergone BH---BH mergers early in their
lifetimes and thus have larger masses, but only one of these has
experienced any substantial accretion events.  In all cases accretion
is effectively quenched when the host galaxies are torn apart and the
BHs are left to wander throughout the halo, and thus any BHs present
here are at or within a few factors of their original seed mass.

\begin{figure} [htb]
\begin{center}
\plotone{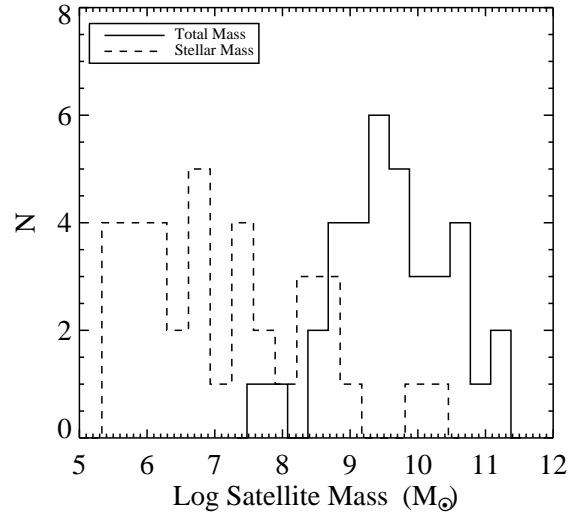}
\epsscale{0.6}
\caption{ The mass distribution of BH-hosting satellites before they
are stripped by the primary.  The solid line indicates the total mass
of the satellites, while the dashed line is the stellar mass only.  Our
simulations are incomplete for halos below a total mass of $3 \times
10^8$ \msun, however due to the UV background field \citep{Haardt96}
most halos with mass below $10^8$ \msun ~have very few stars.
\label{fig:maxmass}}

\end{center} 
\end{figure}

Local Group dwarf galaxies are promising targets for IMBH searches
\citep{VanWassenhove10}. In Figure \ref{fig:maxmass} we show the
masses of BH-hosting satellite galaxies before the stripping process
has begun.  This distribution peaks between $10^9 - 10^{10}$ \msun; at
halo masses of $3 \times 10^8$ \msun ~the satellites become
increasingly dark.  Since very few stars form in these halos, it
is unlikely that we are underestimating the number hosting massive BH
seeds, given our scenario.  We estimate the absolute magnitudes of
the satellites in these mass ranges using the STARBURST99
\citep{Starburst99} code, using stellar ages and metallicities from
our simulations.  This mass range corresponds to galaxies with
magnitudes fainter than -15 in the $V$ band, which includes several
Local Group dwarfs, including the Large and Small Magellanic Clouds.
Examining these local dwarfs for IMBHs may help constrain the
locations and masses of SMBH seeds.  A confirmed detection of an IMBH
would provide an upper limit to the initial mass of SMBH seeds and
possibly allow us to differentiate between the various proposed
formation mechanisms of such seeds.

\subsection{Connection with Off-Nuclear ULXs?}

While wandering BHs may be present in the Milky Way halo, observing
such objects would require either a fortuitous gravitational lensing
detection, or a triggered accretion event, resulting in X-ray
emission.  We investigate situations where a wandering BH could
undergo substantial accretion in a galactic halo, which could explain
the existence of objects such as HLX-1.

\begin{figure}[htb]
\begin{center}

\plotone{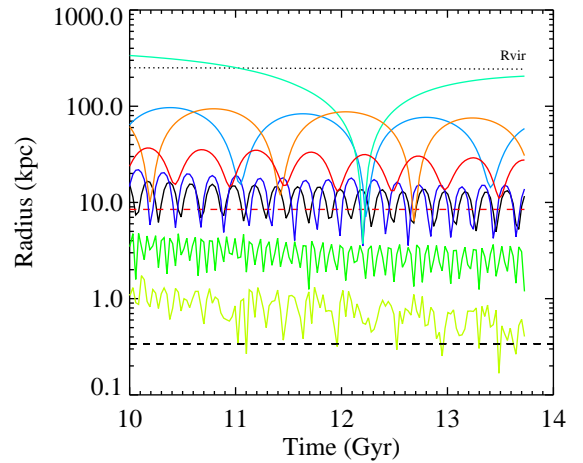}

\caption{ Each curve represents the radial distance vs. time of a
black hole to the halo center for the simulated galaxy h239.  The red
dashed horizontal line denotes a distance of three scale lengths ($r_s
= 2.8$ kpc) from the center; each instance of an orbit crossing this
line is defined as a disk passage.  The black dashed horizontal line
shows the gravitational softening, while the dotted line represents the
virial radius.  The yellow line indicates the central BH, while the
green line denotes an inspiraling BH which has not yet reached the
halo center.  The other curves represent BHs which reside
predominantly in the halo.
\label{fig:orbit}}

\end{center}
\end{figure}

While the majority of ULXs are attributed to star formation regions
\citep{Swartz04}, a smaller number of off-planar ULXs are not
cospatial with star formation, and we focus on the possible origins of
these objects here.  We first examine whether a wandering BH passing
through a dense portion of the host galaxy disk could reproduce the
properties of off-nuclear ULXs.  We trace the orbits of each BH and
examine the frequency with which they cross the disk region, which we
define as three disk scale lengths ($r_s$).  The last few Gyr of
orbital history for the BHs in our simulated galaxy h239 are shown in
Figure \ref{fig:orbit}.  Several of the BHs pass through the $3r_s$
limit, which we denote as a disk passage.  The average BH disk passage
rate for all of the simulations combined is 10.6 per Gyr; however,
this rate increases with time (13.3 per Gyr for the last few billion
years of galaxy evolution) due to the increased population of
wandering BHs in the halo (since more mergers have occurred).

The likelihood of observing such a disk passage depends on the column
density of gas in the disk, the number of passages made, and the
velocity of the BHs.  We use results from HI measurements of the
surface densities of local disk galaxies \citep{Leroy08} to estimate
an average disk column density of 10 \msun ~pc$^{-2}$.  The ability of
the passing BH to attract a sufficient amount of gas to be observed as
a ULX depends most strongly on its velocity, which ranges between 154
- 662 km s$^{-1}$, and on average is $\sim$ 470 km s$^{-1}$ for BHs at
pericenter.  To estimate a BH's luminosity, we developed a simple
scenario where the passing BH is able to collect the gas within a
radius of influence determined by its mass and pericenter velocity:
$M_{coll} = \Sigma_{disk} \pi (GM_{BH}/v_{BH}^2)^2$.  We perform a
Monte Carlo simulation taking into account the observed ranges of BH
velocities, and estimate that the amount of mass collected by a BH
during its disk passage ranges from $10^{-3} - 10^{-4}$ \msun.
Assuming an accretion luminosity $L \sim \eta \dot M c^2$ with $\eta =
0.1$ and the canonical luminosity of $L_{ULX} = 10^{39}$ ergs
s$^{-1}$, the BH will accrete and radiate in a high state for a mean
duration of 2000 years, though a slower BH may radiate for up to
$10^5$ years. However, if the ULX transitions from a high state to a
low state at any point, the accretion rate would drastically decrease
and the observed emission could continue for much longer.

This simple scenario, however, is not sufficient to reproduce the
properties of HLX-1, which is at least 1 kpc away from the plane of
its host galaxy.  A BH passing through the disk moving at a few
hundred km s$^{-1}$ will only travel a distance of $\sim$ tens of
parsecs during the duration of the predicted accretion event.  Such an
event would appear to be a disk ULX source, possibly indistinguishable
from those cospatial with star formation regions.  HLX-1 is also 1000
times more luminous and 100 times less massive (Wiersema et al. 2010.,
in prep.) than the assumed values in our calculation above.  Taking
all of these factors into account, it is unlikely that an isolated
wandering BH passing through a galaxy disk can reproduce the
properties of off-nuclear ULXs.

However, our wandering BH scenario would be feasible if HLX-1 is not
an isolated BH, but is traveling with a bound clump of gas and stars.
The magnitude and colors of the detected optical counterpart of HLX-1
are consistent with a globular cluster \citep{Soria09}.  This cluster
may in fact be the stellar remnants of a stripped dwarf galaxy core
that is still bound to the BH.  HLX-1 could be similar to the object
G1 in M31, a globular cluster that may actually be the nucleus of a
stripped dwarf galaxy \citep{Meylan01} and harbors the most promising
Local Group candidate for an IMBH \citep{Ulvestad07}.  If we presume
that the wandering BHs in our simulations retain a nuclear star
cluster and gas reservoir from their parent halos, then the instance
of a wandering BH passing near the center of the primary could cause
instabilities in its accretion disk, triggering an accretion event of
sufficient magnitude to power a ULX with a high luminosity as seen in
HLX-1.  Our simulations do not have sufficient resolution to follow
the tidally stripped cores of galaxies in detail, though other
simulations have shown that a tidally stripped dwarf galaxy can retain
its core after a close passage with the primary \citep{Mayer03}.
Thus, in the likely instance that a wandering BH retains the core of
its host galaxy, its passage near the galaxy disk can explain the
origin and properties of HLX-1.

 Previous studies have estimated luminosities for massive BHs
wandering through the ISM, but prior to this Letter none have explored
the issue in a cosmological context.  \citet{Krolik04} showed that
IMBHs with masses ranging from $10^2 - 10^4$ \msun ~can produce
luminosities of ULXs if they pass through or near molecular clouds.
SMBH gravitational recoil events passing through the disk may exhibit
X-ray emission of $L > 10^{39}$ ergs s$^{-1}$ \citep{Fujita09}.
\citet{Mapelli08} performed an N-body+SPH simulation of IMBHs in a
galaxy merger, and found that a few halo IMBHs reside in orbits that
pass through the disk, which in our scenario may be observable as
ULXs.

\section{Summary}

We provide compelling dynamical scenario for the presence of
``wandering'' massive BHs in the halos of galaxies. A natural
consequence of the hierarchical build up of galaxies in a $\Lambda$CDM
scenario, the tidal stripping of galaxies containing seed BHs can
populate the halo of a massive disk galaxy with wandering BHs.  These
objects often retain their original seed mass, are found throughout
the galaxy halo, and may pass through the galactic disk at an average
rate of 10.6 Gyr$^{-1}$.  We predict that Local Group dwarf galaxies
such as the Magellanic Clouds are likely to host IMBHs. Detections of
these wandering BHs may give an upper limit to the initial mass of BH
seeds, and may allow us to differentiate between the various proposed
formation mechanisms of such seeds.  Our scenario provides a
physically motivated explanation for off-nuclear ULXs as IMBHs which
have been stripped from their host galaxies, if they retain a gas
reservoir/accretion disk that, when dynamically destabilized, might be
funneled toward the black hole and form an accretion disk.

\acknowledgements

JMB gratefully acknowledges the National Science Foundation Graduate
Research Fellowship Program.  FG acknowledges support from HST
GO-1125, NASA ITP NNX08AG84G, and NSF ITR grant PHY-0205413 (also
supporting TRQ).  MV acknowledges support from SAO TM9-0006X and NASA
ATP NNX10AC84G awards.  Simulations were run using computer resources
and technical support from NAS.  The authors thank Charlotte
Christensen, Adrienne Stilp, Amy Kimball and Brant Robertson for
assistance and helpful comments, as well as the referee whose comments
greatly improved the paper.

\end{document}